\documentstyle[prb,aps]{revtex}
\begin{document}

\title{Fermi-Bose Correspondence at Finite Temperature}
\author{Girish S. Setlur}
\address{Department of Physics and Materials Research Laboratory,\\
 University of Illinois at Urbana-Champaign , Urbana Il 61801}
\maketitle

\begin{abstract}
 The correspondence between fermi-sea/bose-condensate displacements and the 
 number-conserving product of two fermi/bose fields is generalised to
 finite temperatures. It is shown that the straightforward generalisation that
 involves making the sea-bosons participate in the thermodynamic
 averaging does not work out. A remedy is found which involves
 treating the sea-displacement bosons at zero temperature while all finite
 temperature effects have to be lumped into the coefficients.
 We also show that all the finite temperature dynamical four-point
 and six-point functions come out correctly as do some of 
 the commutation rules 
 involving the fermion/boson product that go through unscathed.
 It is also shown that this unusual prescription of leaving out the 
 sea-bosons from the thermodynamic averaging does in fact reproduce the correct
 RPA dielectric function at finite temperature. 
 This article is not however, critic-proof.
\end{abstract}

\section{Introduction}

 The attempts of recent preprints have been to construct a nonperturbative 
 theory of interacting fermions/bosons.  In order for this to have 
 practical consequences, we would like to be able to generalise these ideas
 to finite temperatures where one might be able to study 
 important physical phenomena such as phase transitions. The generalisation
 to finite temperatures is not as straightforward as it seems. The obvious
 generalisation that involves making the sea-displacement bosons 
 participate in the thermodynamic averaging while retaining the original
 formula that relates the fermion/boson product to the sea-displacements
 does not work out. This is explained in detail and a remedy is found.
 The remedy involves making the following unusual prescription. The
 sea-displacement bosons are assumed to remain at zero temperature and do
 not participate in the thermodynamic averaging. Rather, the finite 
 temperature effects have to be lumped into the coefficients thereby altering
 the relation that connects the fermion/boson product and sea-displacements.
 It must be made sure that in the bargain, commutaion rules involving the 
 number-conserving products are not altered. All this is shown explicitly
 in the next few sections.

\section{Fermion Product and Sea-Displacement Correspondence}

 To motivate the development of the finite temperature correspondence
 let us first try to argue that the naive generalisation that seems
 very natural does not work out. One would expect from our earlier
 preprint \cite{Setlurfermi} that at finite temperature, the correlation
 functions in the bose language are merely thermodynamic expectation
 values of the corresponding dynamical quantities. This does not
 seem to work out as this simple demonstration shows. 
 Take for example, the momentum distribution at finite temerature,
\[
\langle c^{\dagger}_{ {\bf{k}} }c_{ {\bf{k}} } \rangle
 = n_{F}({\bf{k}})
 + \sum_{ {\bf{q}}_{1} }
 \Lambda_{ {\bf{k}} - {\bf{q}}_{1}/2 }(-{\bf{q}}_{1})
 \langle a^{\dagger}_{ {\bf{k}} - {\bf{q}}_{1}/2 }({\bf{q}}_{1})
 a_{ {\bf{k}} - {\bf{q}}_{1}/2 }({\bf{q}}_{1}) \rangle
\]
\begin{equation}
 - \sum_{ {\bf{q}}_{1} }
 \Lambda_{ {\bf{k}} + {\bf{q}}_{1}/2 }(-{\bf{q}}_{1})
 \langle a^{\dagger}_{ {\bf{k}} + {\bf{q}}_{1}/2 }({\bf{q}}_{1})
 a_{ {\bf{k}} + {\bf{q}}_{1}/2 }({\bf{q}}_{1}) \rangle
\end{equation}
Since,
\begin{equation}
\langle c^{\dagger}_{ {\bf{k}} }c_{ {\bf{k}} } \rangle
 = \frac{ tr(e^{-\beta(H-\mu_{F} N)}c^{\dagger}_{ {\bf{k}} }c_{ {\bf{k}} }) }
{ tr(e^{-\beta(H-\mu_{F} N) }) }
\end{equation}
 Assume that the $ n_{F}({\bf{k}}) $ are all evaluated at zero temperature
 as in the previous preprints. 
 Taken at face value, one is obliged to compute the thermodynamic
 expectation values of the bose occupation probabilities assuming the
 chemical potential for the bosons is zero as 
 the total number of fermions commutes with the sea-displacements
 (this means that we are allowed
 to create and destroy any number of bosons). Now such a calculation
 yields an infinite answer for
 $ \langle c^{\dagger}_{ {\bf{k}} }c_{ {\bf{k}} } \rangle $ as the
 sum over all $ {\bf{q}}_{1} $ diverges (is proportional to the
 volume of the system). This means that we are no longer allowed
 to treat the sea-displacements 
 bosons as participating in the thermodynamic averaging. Rather, they
 merely provide the correct time-evolution of the product of fermi
 fields, the thermodynamic aspect has to be absorbed into the
 coefficients $ n_{F}({\bf{k}}) $ and  $ \Lambda_{ {\bf{k}} }({\bf{q}}) $,
 as we shall soon see. For this, let us repostulate a 
 form for the product of fermi fields and argue why this should
 be as it is. The only guide is that all the finite temperature
 dynamical correlation
 functions involving the number conserving object
 $ c^{\dagger}_{ {\bf{k+q/2}} } c_{ {\bf{k-q/2}} } $ should be correctly
 reproduced. The  commutation rules involving these number
 conserving products should not be damaged in the bargain.
 Let us start with the simplest case namely,
 $ \langle c^{\dagger}_{ {\bf{k+q/2}} } c_{ {\bf{k-q/2}} } \rangle $ .
 We know what the answer should be, that is,
\begin{equation}
 \langle c^{\dagger}_{ {\bf{k+q/2}} } c_{ {\bf{k-q/2}} } \rangle
 = \delta_{ {\bf{q = 0}} } n_{F, \beta}({\bf{k}})
\end{equation}
where,
\begin{equation}
 n_{F, \beta}({\bf{k}}) = [ exp(\beta(\epsilon_{ {\bf{k}} }-\mu_{F})) + 1 ]^{-1}
\end{equation}
For this to happen we have to do the following,
\newline
 (1) Treat the bosons as before assuming that they are always at
 zero temperature and with zero chemical potential.
 \newline
 (2) Assume that all finite temperature effects are lumped into
 the coefficients.
\newline
 (3) Fix the coefficients as before by demanding that the finite temperature
 dynamical moments of the number-conserving products of fermi fields
 come out right.  
 In order to cut a long story short, let us merely postulate the form
 of the product and later on argue why this should be as it is.
 The following formulas are true ONLY for temperatures above zero.
 At temperatures exactly equal to zero the answers have already been 
 given elsewhere \cite{Setlurfermi}. For $ {\bf{q}} \neq 0 $ we have,
\[
c^{\dagger}_{ {\bf{k+q/2}} }c_{ {\bf{k-q/2}} }
 = 
 (\sqrt{ \frac{N}{\langle N \rangle} })
[\Lambda^{\beta}_{ {\bf{k}} }( {\bf{q}} )
 a_{ {\bf{k}} }({\bf{-q}}) + \Lambda^{\beta }_{ {\bf{k}} }( {\bf{-q}} )
a^{\dagger}_{ {\bf{k}} }({\bf{q}}) ]
\]
\[
+ T_{1}({\bf{k}},{\bf{q}})
\sum_{  {\bf{ q_{1} }} \neq 0}
a^{\dagger}_{ {\bf{k+q/2-q_{1}/2}} }( {\bf{q}}_{1} )
a_{ {\bf{k-q_{1}/2}} }( {\bf{q}}_{1}-{\bf{q}} )
\]
\begin{equation}
- T_{2}({\bf{k}},{\bf{q}})
\sum_{  {\bf{ q_{1} }} \neq 0}
a^{\dagger}_{ {\bf{k-q/2+q_{1}/2}} }( {\bf{q}}_{1} )
a_{ {\bf{k+q_{1}/2}} }( {\bf{q}}_{1}-{\bf{q}} )
\label{SEA}
\end{equation}
Here,
\[
T_{1}({\bf{k}},{\bf{q}}) = \sqrt{ (1-n_{F,\beta}({\bf{k+q/2}})) }
\sqrt{ (1-n_{F,\beta}({\bf{k-q/2}})) }
\]
\[
T_{2}({\bf{k}},{\bf{q}}) = 
\sqrt{ n_{F,\beta}({\bf{k+q/2}})n_{F,\beta}({\bf{k-q/2}}) }
\]
\begin{equation}
\Lambda^{\beta}_{ {\bf{k}} }({\bf{q}}) = 
\sqrt{ n_{F,\beta}({\bf{k+q/2}})(1 - n_{F,\beta}({\bf{k-q/2}}) ) }
\end{equation}
 It can be shown quite easily that all the dynamical four and six-point
 functions of the object 
 $ c^{\dagger}_{ {\bf{k+q/2}} }c_{ {\bf{k-q/2}} } $
 are recovered exactly(See Appendix).
 For $ {\bf{q}} = 0 $ we have to find another expression in order for the
 kinetic energy operator to come out right. For this let us postulate 
 a form and then compute the coefficients.
\[
c^{\dagger}_{ {\bf{k}} }c_{ {\bf{k}} }
 = n_{F, \beta}({\bf{k}})
+ \sum_{  {\bf{ q_{1} }} \neq 0}
S_{1}({\bf{k}},{\bf{q}}_{1})
a^{\dagger}_{ {\bf{k-q_{1}/2}} }( {\bf{q}}_{1} )
a_{ {\bf{k-q_{1}/2}} }( {\bf{q}}_{1} )
\]
\begin{equation}
- \sum_{  {\bf{ q_{1} }} \neq 0}S_{2}({\bf{k}} , {\bf{q}}_{1})
a^{\dagger}_{ {\bf{k+q_{1}/2}} }( {\bf{q}}_{1} )
a_{ {\bf{k+q_{1}/2}} }( {\bf{q}}_{1} )
\label{SEA2}
\end{equation}
 The coefficients $ S_{1} $ and $ S_{2} $ here have to be chosen so as
 to recover the following form of the kinetic energy operator. 
 The latter is necessary to ensure that the time-evolution of the various 
 operators come out right. The correct form of the kinetic energy
 operator is therefore (taking a cue from the zero temperature case
\cite{Setlurfermi}),
\begin{equation}
K = \sum_{ {\bf{k}}, {\bf{q}} }
(\frac{ {\bf{k.q}} }{m})a^{\dagger}_{ {\bf{k}} }({\bf{q}})
a_{ {\bf{k}} }({\bf{q}})
\end{equation}
This means,
\begin{equation}
\epsilon_{ {\bf{k+q/2}} }S_{1}({\bf{k+q/2}}, {\bf{q}})
 - \epsilon_{ {\bf{k-q/2}} }S_{2}({\bf{k-q/2}}, {\bf{q}})
 = (\frac{ {\bf{k.q}} }{m})
\end{equation}
The obvious solution to this is,
\begin{equation}
S_{1}({\bf{k+q/2}}, {\bf{q}}) = S_{2}({\bf{k-q/2}}, {\bf{q}}) = 1
\end{equation}
 Thus we may rewrite the above formula for the occupation number operator as,
\[
c^{\dagger}_{ {\bf{k}} }c_{ {\bf{k}} }
 = n_{F, \beta}({\bf{k}})
+
\sum_{  {\bf{ q_{1} }} \neq 0}
a^{\dagger}_{ {\bf{k-q_{1}/2}} }( {\bf{q}}_{1} )
a_{ {\bf{k-q_{1}/2}} }( {\bf{q}}_{1} )
\]
\begin{equation}
- \sum_{  {\bf{ q_{1} }} \neq 0}
a^{\dagger}_{ {\bf{k+q_{1}/2}} }( {\bf{q}}_{1} )
a_{ {\bf{k+q_{1}/2}} }( {\bf{q}}_{1} )
\label{SEA3}
\end{equation}
 It is straightforward to show that these formulas reproduce the
 correct commutaion rules described below ($ {\bf{q}} \neq 0 $
 and $ {\bf{q}}^{'} \neq 0$) at least in so far as the terms
 linear in the sea-displacements are concerened. The claim is
 that practical results such as the answer for the dynamical current-current
 and density-density correlations should have the right form even though
 many of the nuances may not be right. Think about it, we have a whole
 class of exactly solvable models in any number of spatial dimensions
 and are able to recover RPA and all the other well-known results in terms
 of quadratures, a feature that is likely to persist even after introducing
 more complicated physical processess such as interaction with phonons. 
\begin{equation}
[c^{\dagger}_{ {\bf{k}}^{'} }c_{ {\bf{k}}^{'} },
c^{\dagger}_{ {\bf{k+q/2}} }c_{ {\bf{k-q/2}} }]
 = c^{\dagger}_{ {\bf{k}}^{'} }c_{ {\bf{k-q/2}} }
\delta_{ {\bf{k}}^{'}, {\bf{k+q/2}} }
 - c^{\dagger}_{  {\bf{k+q/2}} }c_{ {\bf{k}}^{'} }
\delta_{ {\bf{k}}^{'}, {\bf{k-q/2}} }
\label{COMM1}
\end{equation}
and,
\begin{equation}
[c^{\dagger}_{ {\bf{k}}+{\bf{q}}/2 }c_{ {\bf{k}}-{\bf{q}}/2  },
c^{\dagger}_{ {\bf{k}}^{'}+{\bf{q}}^{'}/2 }c_{ {\bf{k}}^{'}-{\bf{q}}^{'}/2 }]
 = c^{\dagger}_{ {\bf{k}}+{\bf{q}}/2 }c_{ {\bf{k}}^{'}-{\bf{q}}^{'}/2 }
\delta_{  {\bf{k}}-{\bf{q}}/2, {\bf{k}}^{'}+{\bf{q}}^{'}/2 }
 - c^{\dagger}_{  {\bf{k}}^{'}+{\bf{q}}^{'}/2  }c_{ {\bf{k}}-{\bf{q}}/2 }
\delta_{ {\bf{k}}^{'}-{\bf{q}}^{'}/2, {\bf{k}}+{\bf{q}}/2 }
\label{COMM2}
\end{equation}
\subsection{Finite Temperature RPA}

 In this subsection, we demonstrate that the RPA dielectric function
 is recovered exactly by selectively
 retaining parts of the coulomb interaction that lead to RPA.
 We know that the kinetic energy in the bose language is given by,
\begin{equation}
H_{kin} = \sum_{ {\bf{k}}, {\bf{q}} }
(\frac{ {\bf{k.q}} }{m})
a^{\dagger}_{ {\bf{k}} }({\bf{q}})a_{ {\bf{k}} }({\bf{q}})
\end{equation}
 For this let us choose,
\begin{equation}
H_{I} = \sum_{ {\bf{q}} \neq 0 }\frac{ v_{ {\bf{q}} } }{2V}
{\tilde{\rho}}_{ {\bf{q}} }{\tilde{\rho}}_{ -{\bf{q}} }
\end{equation}
 where,
\begin{equation}
{\tilde{\rho}}_{ {\bf{q}} } = \sum_{ {\bf{k}} }
[ \Lambda^{\beta}_{ {\bf{k}} }({\bf{q}})
a_{ {\bf{k}} }(-{\bf{q}})
+ \Lambda^{\beta}_{ {\bf{k}} }(-{\bf{q}})
a^{\dagger}_{ {\bf{k}} }({\bf{q}}) ]
\end{equation}
 From this it may be shown that the RPA dielectric function
 is recovered as the following demonstration shows. Assume that
 a weak time-varying external perturbation is applied as shown below,
\begin{equation}
H_{ext} = \sum_{ {\bf{q}} \neq 0 }
(U_{ext}({\bf{q}},t) + U^{*}_{ext}(-{\bf{q}},t))
{\tilde{\rho}}_{ {\bf{q}} }
\end{equation}
here,
\begin{equation}
 U_{ext}({\vec{r}},t) = U_{0}
\mbox{  }e^{i{\bf{q}}.{\vec{r}} -i \omega \mbox{ }t}
\end{equation}
Let us now write down the equations of motion for the varios bose fields,
\[
i\frac{ \partial }{\partial t} \langle a^{t}_{ {\bf{k}} }({\bf{q}}) \rangle
 = \omega_{ {\bf{k}} }({\bf{q}})
\langle a^{t}_{ {\bf{k}} }({\bf{q}})  \rangle
 + (\frac{ v_{ {\bf{q}} } }{V})\Lambda^{\beta}_{ {\bf{k}} }(-{\bf{q}})
\sum_{ {\bf{k}}^{'} }[\Lambda^{\beta}_{ {\bf{k}}^{'} }(-{\bf{q}})
\langle a^{t}_{ {\bf{k}}^{'} }({\bf{q}})  \rangle
+ \Lambda^{\beta}_{ {\bf{k}}^{'} }({\bf{q}})
\langle a^{t\dagger}_{ {\bf{k}}^{'} }(-{\bf{q}})  \rangle ]
\]
\begin{equation}
+ (U_{ext}({\bf{q}},t)+ U^{*}_{ext}(-{\bf{q}},t))
\Lambda^{\beta}_{ {\bf{k}} }(-{\bf{q}})
\end{equation}
\[
-i\frac{ \partial }{\partial t}
\langle a^{t\dagger}_{ {\bf{k}} }(-{\bf{q}}) \rangle
 = \omega_{ {\bf{k}} }(-{\bf{q}})
\langle a^{t\dagger}_{ {\bf{k}} }(-{\bf{q}})  \rangle
 + (\frac{ v_{ {\bf{q}} } }{V})\Lambda^{\beta}_{ {\bf{k}} }({\bf{q}})
\sum_{ {\bf{k}}^{'} }[\Lambda^{\beta}_{ {\bf{k}}^{'} }(-{\bf{q}})
\langle a^{t}_{ {\bf{k}}^{'} }({\bf{q}})  \rangle
+ \Lambda^{\beta}_{ {\bf{k}}^{'} }({\bf{q}})
\langle a^{t\dagger}_{ {\bf{k}}^{'} }(-{\bf{q}})  \rangle ]
\]
\begin{equation}
+ (U_{ext}({\bf{q}},t)+ U^{*}_{ext}(-{\bf{q}},t))
\Lambda^{\beta}_{ {\bf{k}} }({\bf{q}})
\end{equation}
Now let us decompose the expectation values as follows,
\begin{equation}
\langle a^{t}_{ {\bf{k}} }({\bf{q}})  \rangle
 = U_{ext}({\bf{q}},t)C_{ {\bf{k}} }({\bf{q}})
+ U^{*}_{ext}(-{\bf{q}},t)D_{ {\bf{k}} }({\bf{q}})
\end{equation}
\begin{equation}
\langle a^{t\dagger}_{ {\bf{k}} }(-{\bf{q}})  \rangle
 = U^{*}_{ext}(-{\bf{q}},t)C^{*}_{ {\bf{k}} }(-{\bf{q}})
+ U_{ext}({\bf{q}},t)D^{*}_{ {\bf{k}} }(-{\bf{q}})
\end{equation}
\begin{equation}
\langle a^{t\dagger}_{ {\bf{k}} }(-{\bf{q}})  \rangle
 = U^{*}_{ext}(-{\bf{q}},t)C^{*}_{ {\bf{k}} }(-{\bf{q}})
+ U_{ext}({\bf{q}},t)D^{*}_{ {\bf{k}} }(-{\bf{q}})
\end{equation}
The coefficients $ C_{ {\bf{k}} }({\bf{q}}) $ and
 $ D^{*}_{ {\bf{k}} }(-{\bf{q}}) $ satisfy,
\begin{equation}
\omega \mbox{ }C_{ {\bf{k}} }({\bf{q}}) = \omega_{ {\bf{k}} }({\bf{q}})
C_{ {\bf{k}} }({\bf{q}})
+ (\frac{ v_{ {\bf{q}} } }{V})\Lambda^{\beta}_{ {\bf{k}} }(-{\bf{q}})
\sum_{ {\bf{k}}^{'} }[\Lambda^{\beta}_{ {\bf{k}}^{'} }(-{\bf{q}})
C_{ {\bf{k}}^{'} }({\bf{q}}) + \Lambda^{\beta}_{ {\bf{k}}^{'} }({\bf{q}})
D^{*}_{ {\bf{k}}^{'} }(-{\bf{q}})]
+ \Lambda^{\beta}_{ {\bf{k}} }(-{\bf{q}})
\end{equation}
\begin{equation}
-\omega \mbox{ }D^{*}_{ {\bf{k}} }(-{\bf{q}})
= \omega_{ {\bf{k}} }(-{\bf{q}})
D^{*}_{ {\bf{k}} }(-{\bf{q}})
+ (\frac{ v_{ {\bf{q}} } }{V})\Lambda^{\beta}_{ {\bf{k}} }({\bf{q}})
\sum_{ {\bf{k}}^{'} }[\Lambda^{\beta}_{ {\bf{k}}^{'} }({\bf{q}})
D^{*}_{ {\bf{k}}^{'} }(-{\bf{q}}) + \Lambda^{\beta}_{ {\bf{k}}^{'} }(-{\bf{q}})
C_{ {\bf{k}}^{'} }({\bf{q}})]
+ \Lambda^{\beta}_{ {\bf{k}} }({\bf{q}})
\end{equation}
Now, effective potential may be written as,
\begin{equation}
U_{eff}({\bf{q}},t) = U_{ext}({\bf{q}},t) +
(\frac{ v_{ {\bf{q}} } }{V})\langle \rho_{ -{\bf{q}} } \rangle^{'}
U_{ext}({\bf{q}},t)
\end{equation}
here,
\begin{equation}
\langle \rho_{ -{\bf{q}} } \rangle
 = U_{ext}({\bf{q}},t)\langle \rho_{ -{\bf{q}} } \rangle^{'}
+ U^{*}_{ext}(-{\bf{q}},t)\langle \rho_{ -{\bf{q}} } \rangle^{''}
\end{equation}
Using the fact tthat,
\begin{equation}
\langle \rho_{ -{\bf{q}} } \rangle^{'}
 = \sum_{ {\bf{k}} }\Lambda^{\beta}_{ {\bf{k}} }(-{\bf{q}})
C_{ {\bf{k}} }({\bf{q}}) + \sum_{ {\bf{k}} }
\Lambda^{\beta}_{ {\bf{k}} }({\bf{q}})D^{*}_{ {\bf{k}} }(-{\bf{q}})
\end{equation}
Solving these equations and using the fact that the dielectric function
 is just the ratio of the external divided by the effective
 potential we get,
\begin{equation}
\epsilon( {\bf{q}}, \omega)
 = 1 + \frac{ v_{ {\bf{q}} } }{V}\sum_{ {\bf{k}} }
\frac{ n_{F,\beta}({\bf{k+q/2}}) - n_{F,\beta}({\bf{k-q/2}}) }
{ \omega - \frac{ {\bf{k.q}} }{m} }
\end{equation}
Which is, lo and behold, the RPA of Bohm and Pines.

\section{Bose Product and Condensate-Displacement Correspondence}
 
 The generalisation to bose systems is entirely analogous. The answers will 
 merely be quoted here, in fact it is easier to verify the various assertions in
 the bose case than it is in the fermi case. Again these formulas are
 valid strictly for temperatures above zero. At exctly zero temperature,
 the answers have been given elsewhere\cite{Setlurbose}. 
\[
b^{\dagger}_{ {\bf{k}} }b_{ {\bf{k}} }
 = n_{B,\beta}({\bf{k}}) 
 + \sum_{  {\bf{ q_{1} }} \neq 0}
d^{\dagger}_{ {\bf{k-q_{1}/2}} }( {\bf{q}}_{1} )
d_{ {\bf{k-q_{1}/2}} }( {\bf{q}}_{1} )
\]
\begin{equation}
- \sum_{  {\bf{ q_{1} }} \neq 0}
d^{\dagger}_{ {\bf{k+q_{1}/2}} }( {\bf{q}}_{1} )
d_{ {\bf{k+q_{1}/2}} }( {\bf{q}}_{1} )
\label{SEAB}
\end{equation}
and for $ {\bf{q}} \neq 0 $,
\[
b^{\dagger}_{ {\bf{k+q/2}} }b_{ {\bf{k-q/2}} }
 =
 (\sqrt{ \frac{N}{\langle N \rangle} })
[\Lambda^{B,\beta}_{ {\bf{k}} }( {\bf{q}} )
 d_{ {\bf{k}} }({\bf{-q}}) + \Lambda^{B,\beta }_{ {\bf{k}} }( {\bf{-q}} )
d^{\dagger}_{ {\bf{k}} }({\bf{q}}) ]
\]
\[
+ T_{1}({\bf{k}},{\bf{q}})
\sum_{  {\bf{ q_{1} }} \neq 0}
d^{\dagger}_{ {\bf{k+q/2-q_{1}/2}} }( {\bf{q}}_{1} )
d_{ {\bf{k-q_{1}/2}} }( {\bf{q}}_{1}-{\bf{q}} )
\]
\begin{equation}
+ T_{2}({\bf{k}},{\bf{q}})
\sum_{  {\bf{ q_{1} }} \neq 0}
d^{\dagger}_{ {\bf{k-q/2+q_{1}/2}} }( {\bf{q}}_{1} )
d_{ {\bf{k+q_{1}/2}} }( {\bf{q}}_{1}-{\bf{q}} )
\label{SEAB2}
\end{equation}
and the coefficients are similarly given as,
\begin{equation}
T_{1}({\bf{k}}, {\bf{q}} )
 = \sqrt{ (1 + n_{B,\beta}({\bf{k}}+{\bf{q}}/2)) }
\sqrt{ (1 + n_{B,\beta}({\bf{k}}-{\bf{q}}/2)) }
\end{equation}
\begin{equation}
T_{2}({\bf{k}}, {\bf{q}} )
 = \sqrt{ n_{B,\beta}({\bf{k}}+{\bf{q}}/2) }
\sqrt{ n_{B,\beta}({\bf{k}}-{\bf{q}}/2) }
\end{equation}
and,
\begin{equation}
\Lambda^{\beta,B}_{ {\bf{k}} }({\bf{q}})
 = \sqrt{ n_{B}^{\beta}({\bf{k}}+{\bf{q}}/2)
( 1 + n_{B}^{\beta}({\bf{k}}-{\bf{q}}/2) ) }
\end{equation}
and,
\begin{equation}
n_{B}^{\beta}({\bf{k}}) =
\frac{1}{ e^{ \beta( \epsilon_{ {\bf{k}} }-\mu_{B} ) } - 1 }
\end{equation}
There are two points worth noting here, one is the change of
 sign in $ ( 1 + n_{B}^{\beta}({\bf{k}}-{\bf{q}}/2) ) $ in the fermi
 case it was $ ( 1 - n_{F}^{\beta}({\bf{k}}-{\bf{q}}/2) ) $, the reason
 should be obvious. The other is the overall change in sign in $ T_{2} $,
 this is necessary to ensure that the correct six point function is recovered.
 The proof that all the six-point functions come out right may be seen in
 Appendix.

\section{Appendix}
Assume that in the fermi case the part quadratic in the sea-displacements
 is given by,
\[
c^{\dagger}_{ {\bf{k+q/2}} }c_{ {\bf{k-q/2}} }|_{Quad}
 = \sum_{ {\bf{k}}_{1}, {\bf{k}}_{2}, {\bf{q}}_{1}, {\bf{q}}_{2} }
\Gamma_{ {\bf{k}}_{1}, {\bf{k}}_{2} }^{ {\bf{q}}_{1}, {\bf{q}}_{2} }
({\bf{k}}, {\bf{q}})a^{\dagger}_{ {\bf{k}}_{1} }({\bf{q}}_{1})
a_{ {\bf{k}}_{2} }({\bf{q}}_{2})
\]
From this we may write,
\[
I = \langle c^{\dagger}_{ {\bf{k+q/2}} }c_{ {\bf{k-q/2}} }
c^{\dagger}_{ {\bf{k^{'}+q^{'}/2}} }c_{ {\bf{k^{'}-q^{'}/2}} }
c^{\dagger}_{ {\bf{k^{''}+q^{''}/2}} }c_{ {\bf{k^{''}-q^{''}/2}} } \rangle
\]
\[
 = [(1 - n_{F, \beta}({\bf{k-q/2}}))(1 - n_{F, \beta}({\bf{k^{'}-q^{'}/2}}))
n_{F, \beta}({\bf{k+q/2}})\delta_{ {\bf{k+q/2}}, {\bf{k^{''} - q^{''}/2}} }
\delta_{ {\bf{k-q/2}}, {\bf{k^{'} + q^{'}/2}} }
\delta_{ {\bf{k^{'}-q^{'}/2}}, {\bf{k^{''} + q^{''}/2}} }
\]
\begin{equation}
 - (1 - n_{F, \beta}({\bf{k-q/2}}))n_{F, \beta}({\bf{k^{'}+q^{'}/2}})
n_{F, \beta}({\bf{k+q/2}})
\delta_{ {\bf{k-q/2}}, {\bf{k^{''} + q^{''}/2}} }
\delta_{ {\bf{k+q/2}}, {\bf{k^{'} - q^{'}/2}} }
\delta_{ {\bf{k^{'}+q^{'}/2}}, {\bf{k^{''} - q^{''}/2}} } ]
\end{equation}
In terms of the Bose fields we have,
\begin{equation}
I = \Lambda^{\beta}_{ {\bf{k}} }({\bf{q}})
\Lambda^{\beta}_{ {\bf{k}}^{''} }(-{\bf{q}}^{''})
 \Gamma_{ {\bf{k}}, {\bf{k}}^{''} }^{ -{\bf{q}}, {\bf{q}}^{''} }
({\bf{k}}^{'}, {\bf{q}}^{'})
\end{equation}
This leads to the answers for $ T_{1} $  and $ T_{2} $ given in the main
 text. The arguments in the bose case is analogous.

\section{ACKNOWLEDGEMENTS}
It is a pleasure to thank Prof. A. H. Castro-Neto and Prof. D. K. Campbell
 for providing important references and encouragement and
 Prof. A. J. Leggett for giving his valuable time and advice
 on matters related to the pursuit of this work. Thanks are also due to
 Prof. Ilias E. Perakis for providing the author with an important
 reference and to Prof. Y.C. Chang for general encouragement.
 This work was supported in part by ONR N00014-90-J-1267 and
 the Unversity of Illinois, Materials Research Laboratory under grant
 NSF/DMR-89-20539 and in part by the Dept. of Physics at
 University of Illinois at Urbana-Champaign. The author may be contacted at
 the e-mail address setlur@mrlxpa.mrl.uiuc.edu.


\newpage
\begin{thebibliography}{3}

\bibitem[1]{Setlurfermi}  G. S. Setlur,
"Expressing Products of Fermi Fields in terms of Fermi-Sea displacements",
 UIUC preprint 1997, cond-mat/9701206, and other unpublished formulae.
"Exact Momentum Distribution of a Fermi Gas in One Dimension",
 UIUC preprint 1997. cond-mat/9705219

\bibitem[2]{Setlurbose}  G. S. Setlur,
 "Exact Dynamical Structure Factor of a Bose Liquid", UIUC preprint 1997.
 cond-mat/9704222

\bibitem[3]{Setlurbosefermi}  G. S. Setlur,
 "Exact Single-Particle Green Functions of Bose and Fermi Liquids", 
 UIUC preprint 1997. hep-th/9706006
                                  

\end{thebibliography}
\end{document}